# Mo-Si-B alloys for ultra-high temperature space and ground applications: liquid assisted fabrication under various temperature and time conditions

Grzegorz Bruzda[1,2], Wojciech Polkowski[1*], Adelajda Polkowska[1], Rafał Nowak[1], Artur Kudyba[1], Marzanna Książek[2], Sebastian Lech[3], Krzysztof Karczewski[4], Donatella Giuranno[5]

[1]*Łukasiewicz Research Network – Krakow Institute of Technology, Zakopiańska 73 Str., 30-418 Cracow, Poland*
[2]*AGH University of Science and Technology, Faculty of Non-Ferrous Metals, Mickiewicza 30 Av., 30-059 Krakow, Poland*
[3]*AGH University of Science and Technology, International Centre of Electron Microscopy for Materials Science and Faculty of Metals Engineering and Industrial Computer Science, Mickiewicza 30 Av., 30-059 Krakow, Poland*
[4]*Faculty of Advanced Technologies and Chemistry, Military University of Technology, gen. S. Kaliskiego 2 Str., 01-476 Warsaw, Poland*
[5]*CNR - Institute of Condensed Matter Chemistry and Technologies for Energy, Via E. De Marini, 6- 16149 Genova, Italy*
*Corresponding author: grzegorz.bruzda@kit.lukasiewicz.gov.pl; wojciech.polkowski@kit.lukasiewicz.gov.pl;*

**ORCID identifiers:**

Grzegorz Bruzda: 0000-0002-6965-1958

Wojciech Polkowski: 0000-0003-2662-7824

Adelajda Polkowska: 0000-0002-7617-1098

Rafał Nowak: 0000-0001-8096-5176

Artur Kudyba: 0000-0003-3065-7778

Marzanna Książek: 0000-0001-6377-9363

Sebastian Lech: 0000-0001-6510-1185

Krzysztof Karczewski: 0000-0001-9822-8799

Donatella Giuranno: 0000-0002-7241-5032

**Abstract**

Boron-doped molybdenum silicides have been already recognized as attractive candidates for space and ground ultra-high temperature applications far beyond limits of state-of-the-art nickel based superalloys. In this work, we are exploring a new method for fabricating Mo-Si-B alloys (as coatings or small bulk components) by utilizing a pressure-less reactive melt infiltration approach. The basic assumption of this approach is a synthesis of binary and/or ternary and complex intermetallic phases (silicides, borides, borosilicides), through a direct interaction of Si-B melt with molybdenum . The main purpose of this work, was to examine the effect of temperature and time of Si-B melt interaction on the structure and morphology of the formed reaction products. For this purpose, sessile drop experiments were carried out on the eutectic Si-3.2B (wt%) alloy/Mo couples at temperature varying between 1385-1550°C and holding time between 10 to 30 minutes. The solidified sessile drop couples were subjected to microstructural characterization by means of light microscopy and scanning electron microscopy analyses performed both at "top-view" and cross-sectioned interfaces. The phases formed within the interaction zone were identified by using TEM/SAED and XRD techniques. It was documented that a thickness of both main product layer ($MoSi_2+Mo_5Si_3$), as well as boron-rich interlayer increases with raising temperature and time of the Si-B melt interaction with Mo substrates.

## 1. Introduction

The only way for making real the slogan "*The hotter engine the better*" is to introduce new structural materials with improved high temperature strength and durability. As actually applied metallic heat resistant materials, namely Ni-, Co- or Fe- based superalloys have reached their maximum working temperatures (as it comes from their melting points and insufficient high temperature strength), research focuses and efforts have been shifted towards advanced ceramic composites or refractory metals (RM) based alloys [1]. Thus, in order to produce and use energy in a more efficient and responsible manner, the introduction of new materials "beyond superalloys" (i.e. able to operate at temperatures above ~1150°C), is mandatory.

Pure refractory metals, due to intrinsic drawbacks (i.e. very low oxidation resistance, high specific weight and low mechanical strength) are not suitable for replacing superalloys in energy or aviation sectors [2]. In this regard, much better properties have been documented for RM based intermetallics like Mo or Nb-silicides. These materials exhibit melting points above 2000°C, and a winning combination of high strength and low density (6-7 gcm$^{-3}$ vs. 7-8 gcm$^{-3}$ of superalloys) [3]. Furthermore, as simple binary silicides suffer from the so-called "pest oxidation" phenomenon at intermediate temperatures (i.e. rapid degradation due to a lack of continuous protective surface layer), this problem has been already solved by microalloying with B (1-3 wt%). The modification of the chemical composition provides a significant improvement of the oxidation resistance by promoting the growth of a borosilicate glass layer at the surface showing a low viscosity and a "self-healing" behavior [4]. Moreover, the alloying with B induces new binary (borides) or ternary (borosilicides) phases having high hardness and creep resistance. Consequently, it has been also proven that B improves the performance of Mo-based silicides airfoils for in aircraft engines, by increasing the usual working temperature of Ni-based single-crystalline superalloys of ~150°C [5], .

Despite such outstanding and proved improved performance, intrinsic properties of Mo-Si-B alloys still limit the use of "classical" fabrication methods. Recently reported fabrication procedures for Mo-Si-B alloys consist in multi-stage processes in nature and involve either very high temperature/pressure conditions (costly powder metallurgy-based techniques) [5] or the presence of toxic halide activators highly detrimental for both environment and human health (i.e. pack cementation process) [4].

In our previous work [6], encouraging results about the feasibility of using a new clean and cost-effective approach for the fabrication of B alloyed Mo silicides based materials have been reported. The main idea is to utilize the Reactive Melt Infiltration (RMI) approach to synthesize multiphase Mo-Si-B materials. In particular, the RMI process is based on a reactive interaction of Si-B melts with Mo substrates/templates under certain temperature/pressure/time conditions. The RMI method itself has been already commercialized, and the most spectacular example of its application is the large-scale fabrication of near-net shaped SiC/C ceramic matrix composites for aviation and space applications [7]. Here, the concept of liquid-assisted siliconizing process proposed by Zhang *et al.* [8-16] for the fabrication of Si–MoSi$_2$ functionally graded coatings or bulk materials by a hot-dipping of Mo substrate in a Si bath, is adopted. This method offers many advantages over other ones, such as a simple experimental setup, pressure-less and clean conditions (no toxic chemical activators), limited temperature ranges, fast diffusion rate of Si, good product quality, reproducibility, etc. The authors documented that functionally graded MoSi$_2$-Mo coatings produced by the proposed method show an excellent oxidation resistance and a good bonding with Mo substrates. They also observed that the morphology of MoSi$_2$-Mo products might be also easily controlled by adjusting the hot-dipping process parameters [17, 18].

The main difference between the reported Zhang's works and the method and results described by the authors, relies in producing *boron enhanced* silicides by a novel approach, based on RMI process, as well as in introducing B directly from Si-B melts. Therefore, the main purpose of this work, was to examine the effect of process parameters (e.g. temperature and time) of a Si-B melt interaction on the formed reaction products at the liquid Si-B/Mo interface and on the developed microstructure and morphology, with a focus on boron enriched phases.

## 2. Materials and methods

Materials used were plates of polycrystalline Mo (99.95 %) with a thickness of 6 mm (Wolften, Poland) and binary eutectic Si-B alloy (Si-3.2B wt%). The selection of Si-B eutectics is justified in terms of a potential lowering of RMI processing temperature ($T_{eSi-3.2B}$=1385°C [19]). Small buttons of Si-3.2B alloy were produced from polycrystalline materials (Si: 99.999%; B: 99.9% provided by Onyxmet, Poland). The main impurities in the batch materials for the fabrication of Si-B alloy were Al (0.008 at.%) and P (0.002 at.%) in silicon and Fe (0.023 at.%) and Al (0.012 at.%) in B. Alloy samples were produced by melting the mixtures in capillaries (at 1500°C/30 min) followed by a melt deposition on ceramic substrates. Both capillaries and substrates were made of h-BN, which shows exceptional inertness towards Si-based melts at high temperature [20-22]. The mentioned process fabrication of alloy samples was carried out in a chamber of our experimental device (Fig. 1a), after producing a preliminary high vacuum conditions ($p=10^{-7}$ mbar), and approaching to the alloy melting an Ar protective atmosphere was introduced for avoiding evaporation phenomena. After the deposition (Fig. 1b, c), the samples were cooled down to room temperature (20°Cmin$^{-1}$) and the check of final composition and microstructure were carried out by SEM/EDS analyses and XRD techniques (Fig. 1d, e).

Before the high temperature experiments, the surface of Mo substrate was gently ground with SiC papers, polished with diamond suspensions and then degreased with isopropanol alcohol.

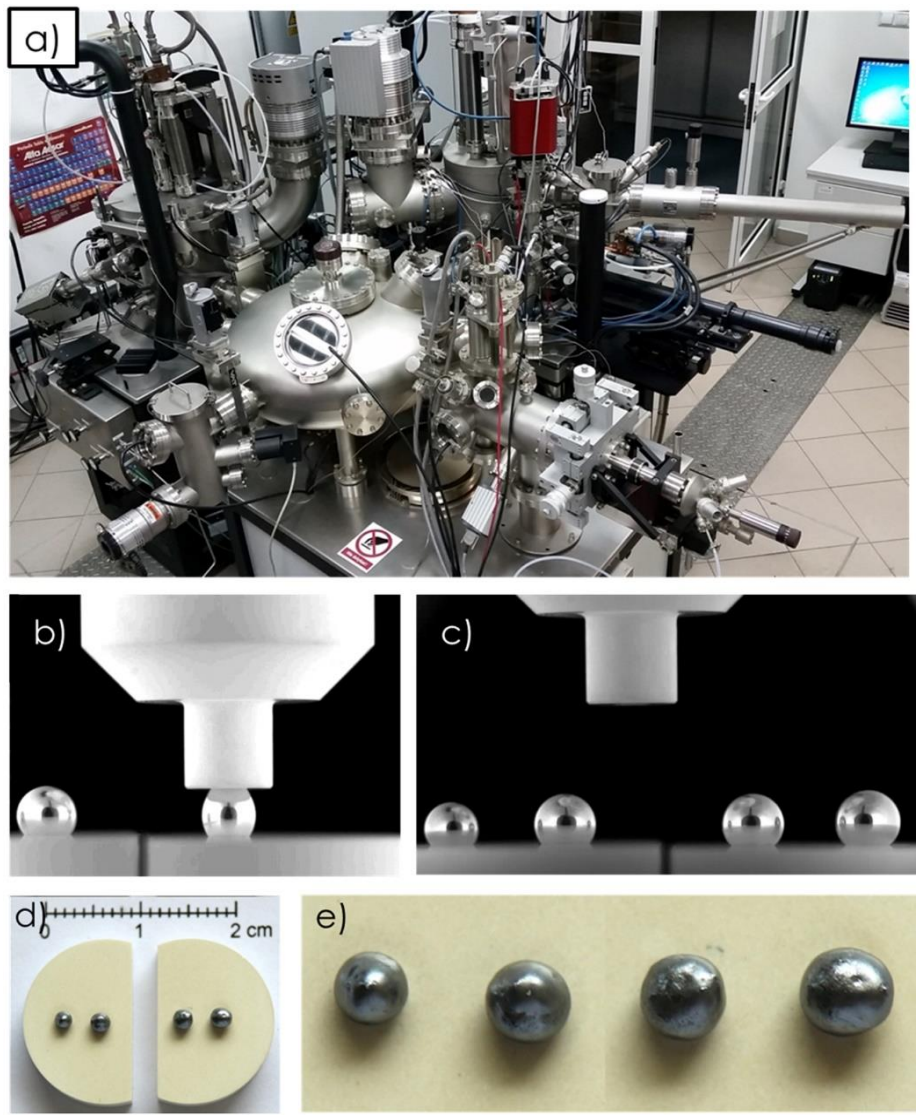

*Fig. 1. An experimental high-temperature device of Łukasiewicz-Krakow Institute of Technology [19](a); a fabrication of Si-3.2B buttons alloy by melting in h-BN capillary followed by a deposition on h-BN substrate (b, c); produced Si-3.2B alloy pieces (d, e).*

The experiments were carried out by using an experimental setup located at Łukasiewicz – Krakow Institute of Technology laboratories (Fig. 1a). It consists of a complex device ad-hoc designed for examining liquid metals at high temperatures, as described in details elsewhere [23]. The experiments were performed by the sessile drop method coupled with the contact heating procedure under well-defined experimental conditions. For each single test a piece of Si-3.2B alloy was placed on the Mo substrate and then the couple was subjected to various heating/cooling runs (Fig. 2).

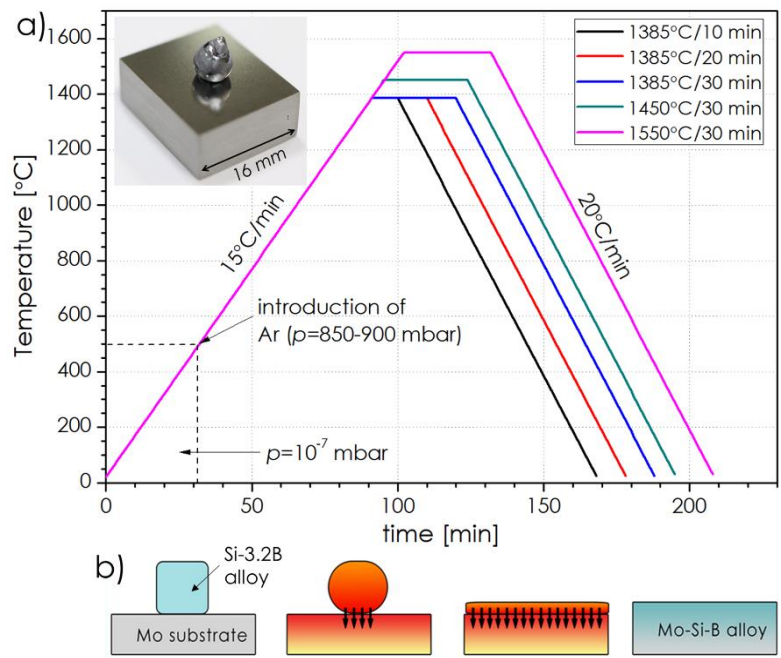

*Fig. 2. Temperature profiles of applied sessile drop experiments (a) and a scheme of the RMI-based process for a fabrication of Mo-Si-B alloys (b). An insert in (a) shows exemplary Si-3.2B/Mo couple before the tests.*

Specifically, the couple was heated up to $T$=500°C (15°Cmin$^{-1}$) under a high vacuum of $p$=10$^{-7}$ mbar; and, static Ar ($p$=850-900 mbar) was introduced into the chamber in order to hinder evaporation of the Si-B alloy. The testing temperature and time were varied between $T$=1385-1550°C and $t$=10-30 min, respectively. After the end of the test, the samples were cooled down to room temperature with a rate of 20 Cmin$^{-1}$. During each experiment, the Si-B alloy/Mo substrate images were in real-time acquired by a high speed camera (up to 100 fps) and off-line used to compile movies and to produce wetting kinetics curves ($\theta$-contact angle vs. time).

The solidified Si-3.2B/Mo couples were subjected to microstructural characterization. The obtained microstructures were analyzed, both at top-view and at the cross-sectioned samples, by using FEI Scios™ field emission gun scanning electron microscope (FEG-SEM) coupled with energy-dispersive X-ray spectroscopy (EDS). Cross-sections for SEM examinations were prepared by cutting the samples with diamond wheels, mechanical grinding and polishing with a colloidal silica suspension. Furthermore, more detailed structural analyses and phase identifications were performed by using transmission electron microscopy (TEM) and selected area electron diffraction (SAED). For this purpose, lamellae were prepared directly from selected areas of the cross-sectioned samples via the Focused Ion Beam (FIB) technique, using the NEON CrossBeam 40EsB microscope (ZEISS), as shown in Fig. 10. In particular, although the thickness was not uniform within the whole sample, due to different milling rate of multiple phases, a protective Ga$^+$ ion layer was sputtered on the cross-sectioned sample, covering all areas of interest – Mo substrate, interlayer and main layer. Samples were thinned down to about 70 nm. The phase identification was performed by using SAED technique and a post-processing with JEMS software. Additionally, the phase composition of the formed products at the interface, was determined by means of X-ray diffraction (Rigaku Ultima IV, Japan) using Co-K$_\alpha$ radiation, a *2θ* scan range of 15–115°, a scan speed of 1°min$^{-1}$ and a step size of 0.02°.

### 3. Results and discussion

### 3.1. Materials in initial state

The FEG-SEM images showing initial microstructures of batch materials are given in Fig. 3. Mo plates exhibited typical equiaxed grain structure (with a mean grain size of ~18 μm). A near eutectic microstructure resulting in a mixture of Si+SiB$_3$ and Si-rich matrix as revealed by local

FEG-SEM/EDS analyses and in fully agreement with the reported Si-B phase diagram [19], was detected at the cross-sectioned alloy sample.

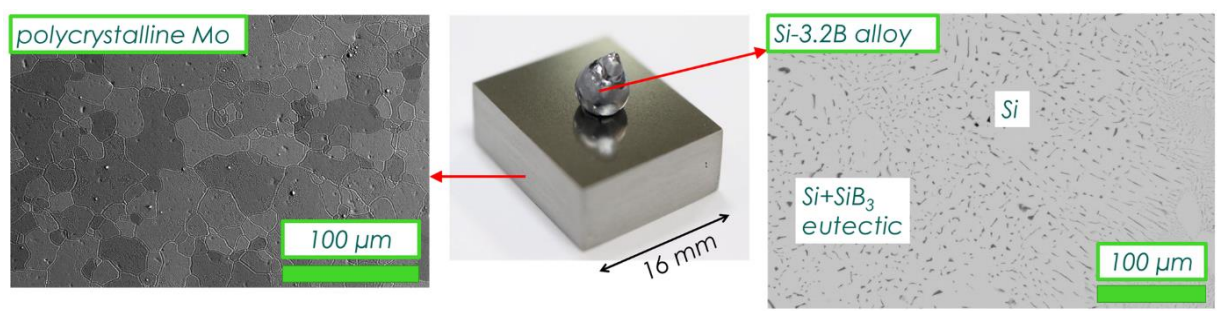

*Fig. 3. FEG SEM images showing typical starting microstructures of polycrystalline Mo and Si-3.2B alloy.*

## 3.2. Wetting and spreading behavior

Selected images of the Si-3.2B alloy/Mo substrate sample recorded during the sessile drop tests are collected in Fig. 4. As it can be observed, the Si-B melt shows a very good wetting in contact with the Mo substrate. Moreover, the wetting kinetics seems to be weakly affected by the imposed testing parameters. In particular, just after the liquid phase appeared, the alloy rapidly spreads over the substrate. Despite the different experimental conditions imposed, the final contact angle value of $\theta$ ~5° was obtained in 60-70 seconds for all wetting experiments performed. As in the case of our previous study [6], the melting started at a temperature between 1345 and 1352°C, i.e. lower than that of the theoretical eutectic temperature for binary Si-3.2B alloy ($T_e$=1385°C). Such discrepancy might be associated to the effect of solid state diffusion affecting local chemical composition in near interface area. Consequently, a ternary eutectic reaction ($L \leftrightarrow (Si)+MoSi_2+SiB_3$) may appear at $T$~1350°C, as it was predicted by Katrych *et al.* [24]. The wetting kinetics plots given in Fig. 5 show that the testing temperature does not significantly affect the wetting and spreading dynamics upon the continuous heating process. On the other hand, it is well known that spreading behaviors of such highly reactive systems, may be affected by the applied contact heating sessile drop method. Indeed, the alloy spreads over the substrate out of thermal equilibrium. It means that the interface is under "dynamic conditions" and its composition may change during the wetting experiment due to the growth of different reaction products occurring during the raise of temperature. In other words, reaching the testing temperature, mainly at temperature higher than melting point, the wetting kinetics may be influenced by transient phenomena already occurred at the interface.

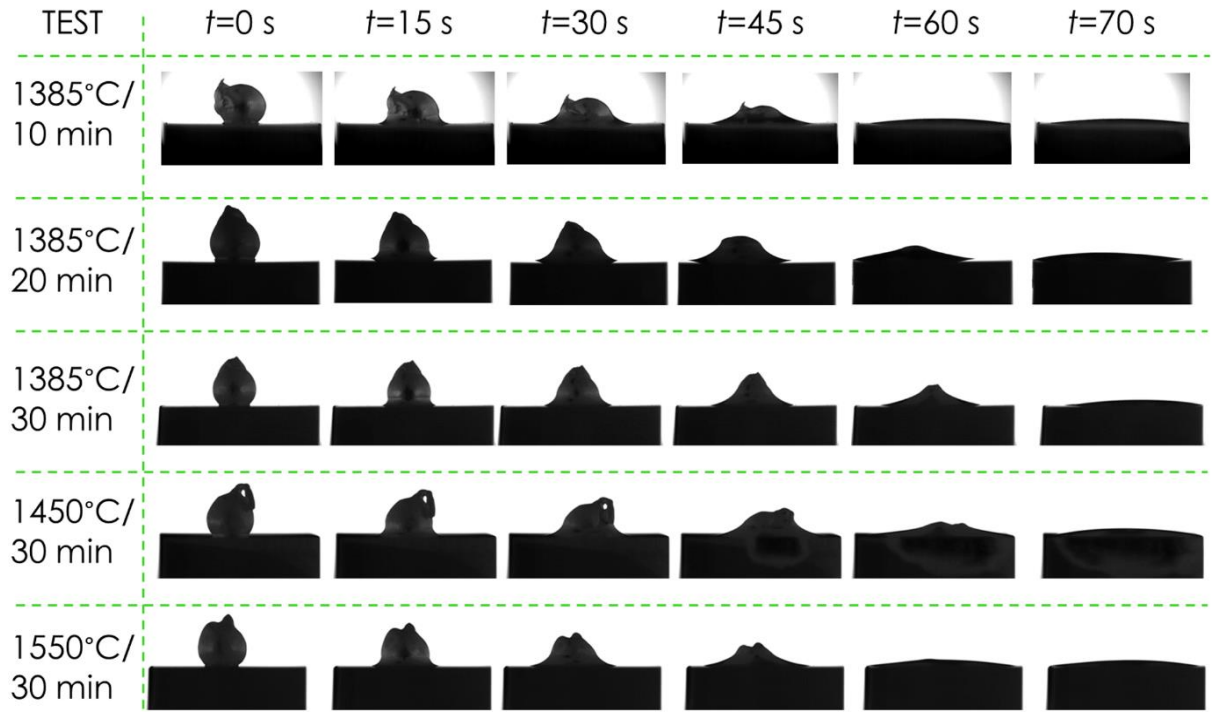

Fig. 4. Selected images recorder by high speed camera upon various sessile drop experiments on Si-3.2B/Mo couples.

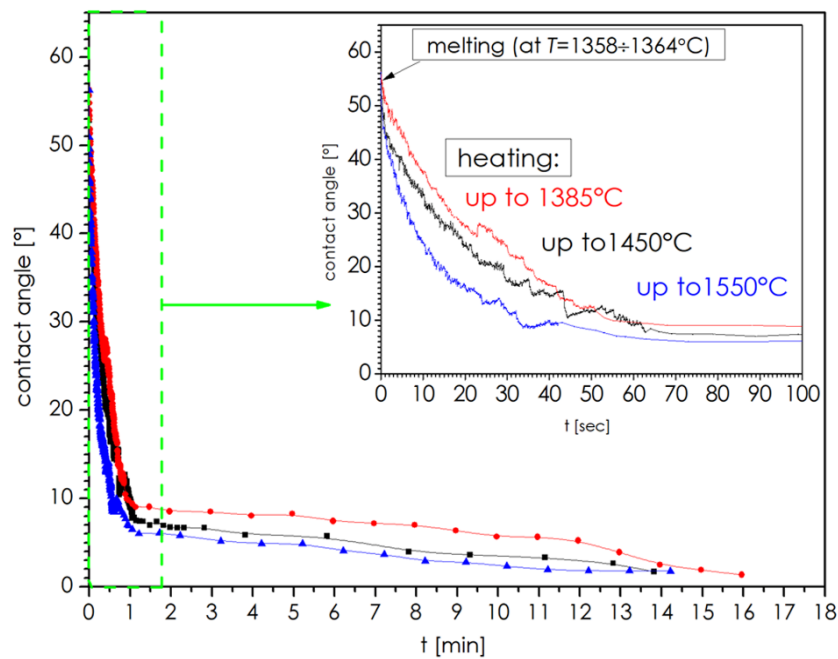

Fig. 5. The wetting kinetics curve (θ vs. t) calculated from acquired drop/substrate images.

### 3.3. Microstructural characterization of solidified Si-3.2B/Mo samples

Top view macro-photos of the solidified couples are shown in Fig. 6. In each case a product layer is easily recognized at the surface of Mo plates. A color changeover (to blueish) on sites non-covered by the liquid-assisted reaction products might be associated with effects related to an evaporation of Si-B melt followed by a condensation on initially polished Mo surfaces. The results of FEG-SEM top-view observations revealed a dense structure of rounded crystals (as observed from the top) and some small amount of retained Si(B) alloy (Fig. 7). The morphology is analogous to that recently reported by Zhang et al. [11] for Mo-MoSi$_2$ composites produced by the hot-dipping approach. Interestingly, the size of crystals seems to be not significantly affected by the applied processing conditions, and in each case it was around 20-40 μm.

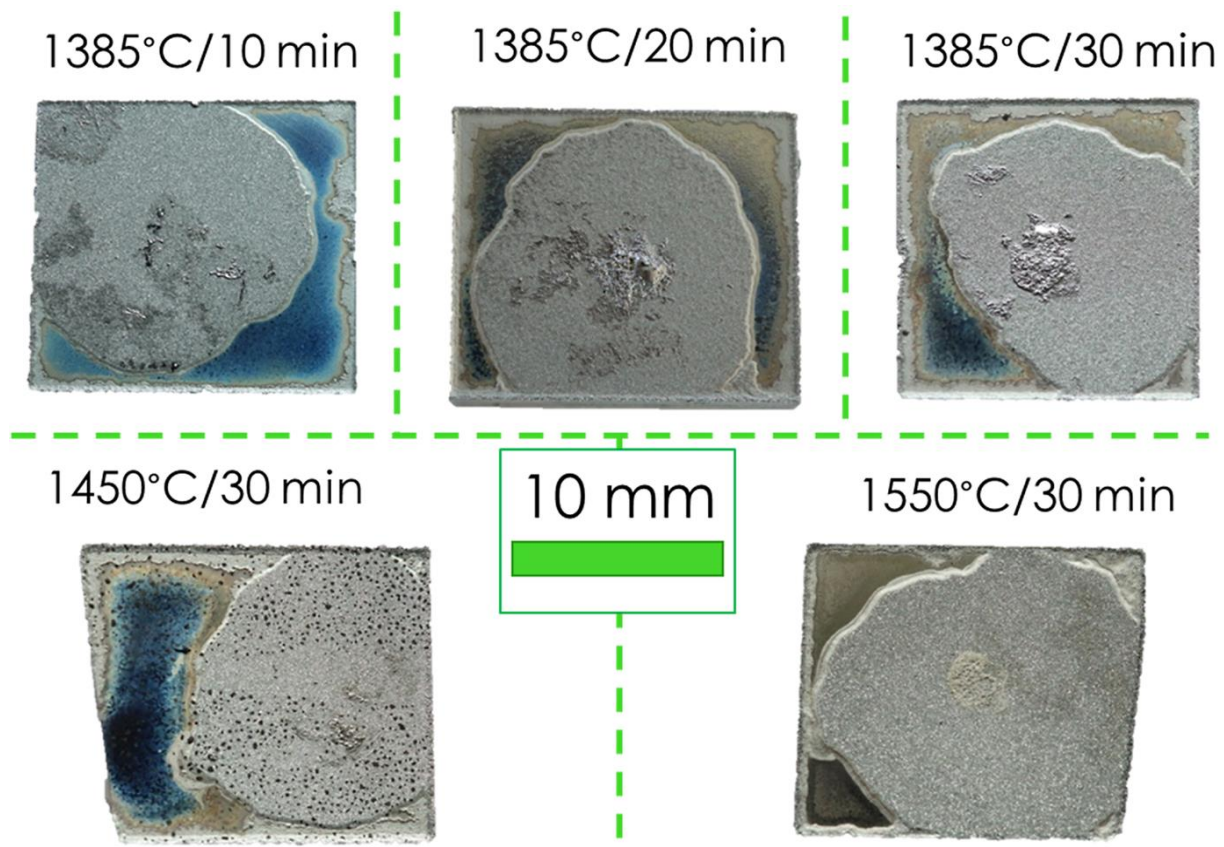

*Fig. 6. Top view macro-images of the Si-3.2B/Mo sessile drop specimens after the tests.*

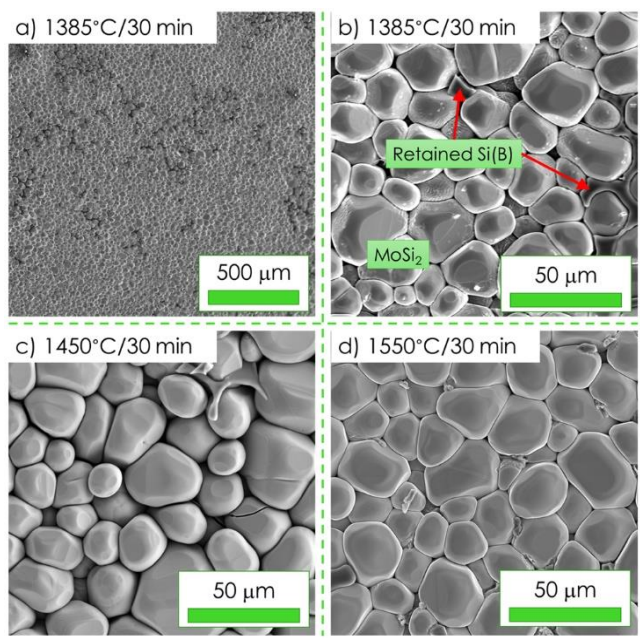

*Fig. 7. Top-view FEG-SEM images of the reactively formed products. A low magnification image showing typical morphology of a dense surface layer (a); high magnification images taken from samples produced at: 1385°C/30 min (b); 1450°C/30 min (c) and 1550°C/30 min (d).*

The results of more detailed SEM/EDS analyses performed on cross-sectioned samples allow revealing that the main product layer consists of $MoSi_2$ columnar crystals with incorporated $Mo_5Si_3$ particles, as shown in Fig. 8. By taking into account the binary Mo-Si phase diagram [25], it might be assumed that both phases were produced in a direct reaction with the Si-rich melt. Additionally, a presence of intermediate continuous layer formed between ($MoSi_2$+$Mo_5Si_3$) products and Mo substrate is clearly recognized (Fig. 8). A thickness of both the main and intermediate layers increased with rising temperature and time of the Si-B melt interaction, reaching up to ~120 and 25 µm, respectively for the sample produced at 1550°C/30 min.

As shown in Figs. 8 and 9, a large number of pores can be found in the outermost layer (main layer). The pores formation has been previously observed in hot dip silicon plating experiments [26], due to a formation of columnar-like crystals inside the reaction area separated by narrow grain boundaries. It has been proposed that under the action of surface tension, molten silicon fails to wet the narrow grain boundaries, which is the main reason for the formation of micro-pores in these sites. Furthermore, another reason for the formation of micro-voides may be explained by so-called Kirkendall effect, that is commonly observed within intermetallic compound layers in solder joints of semiconductor package interconnections [27]. Because of difference in diffusivity of the atoms in a binary system (in here: silicon in the $MoSi_2/Mo_5Si_3$ couple), the mass transfer results in a loss of mass of one component and a mass gain of another one. Subsequently, a vacancy flux passes in the direction having a lower diffusivity, while the vacancy condensation combined with a a state of internal stresses supports the voids formation.

At higher SEM magnifications, three sub-layers were further detected inside the intermediate zone (Fig. 9a). It should be noted that the sub-layers thickness was found to be different at various testing temperatures (Figs. 9b,c). Although the results of SEM/EDS microanalysis point towards a high B-content in intermediate layers, more sophisticated tools were needed to identify individual phases. For this purpose, additional TEM analyses were performed.

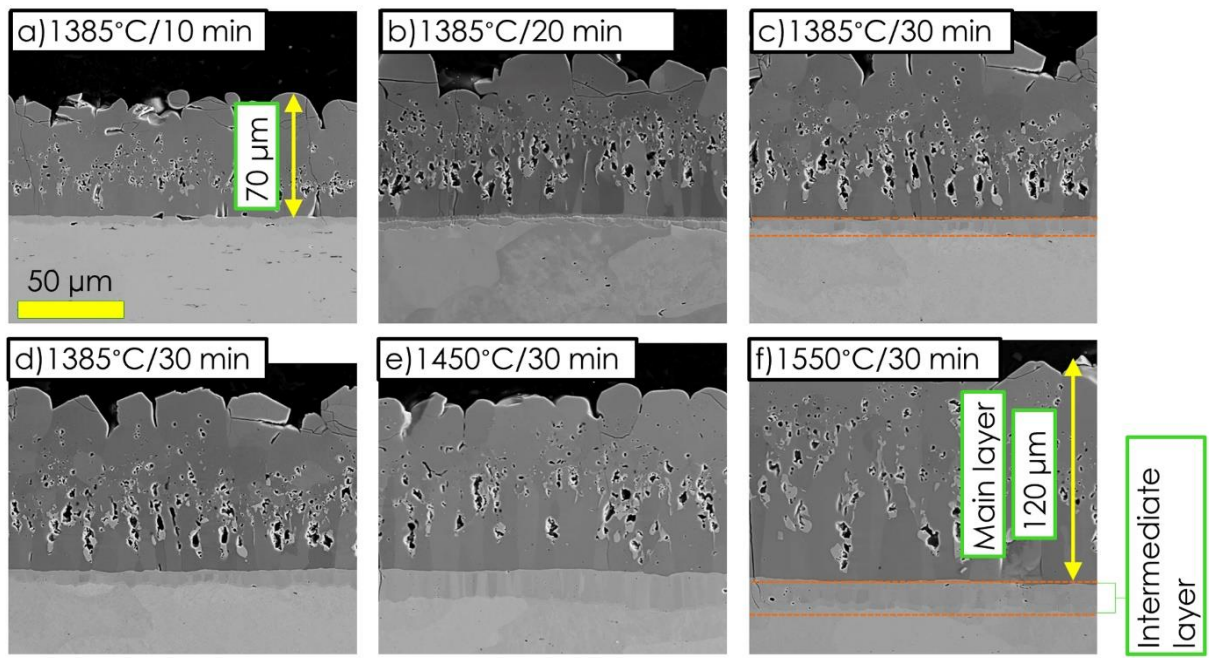

Fig. 8. The results of detailed SEM analyses on cross-sectioned Si-3.2B/Mo sessile drop samples showing a general view of the product layer produced at: 1385°C/10 min (a); 1385°C/20 min (b); 1385°C/30 min (c, d); 1450°C/30 min (e) and 1550°C/30 min (f).

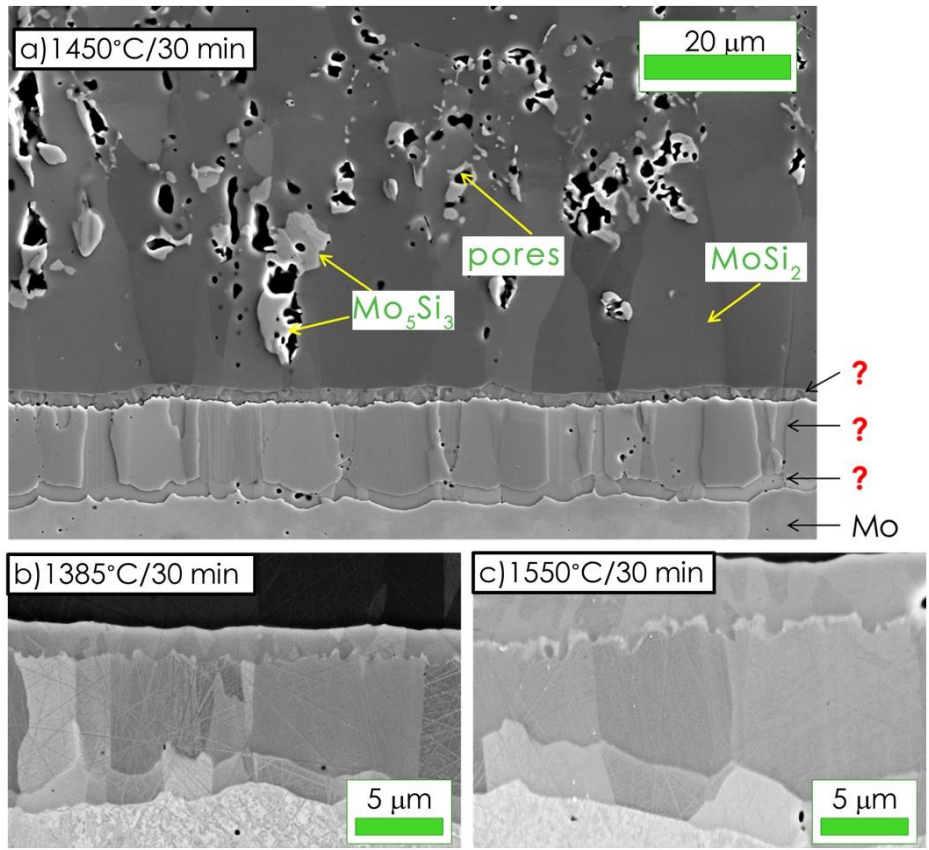

*Fig. 9. High magnification SEM images of cross-sectioned Si-3.2B/Mo sessile drop samples showing the product layer produced at 1450°C/30 min (a); 1385°C/30 min (b) and 1550°C/30 min.*

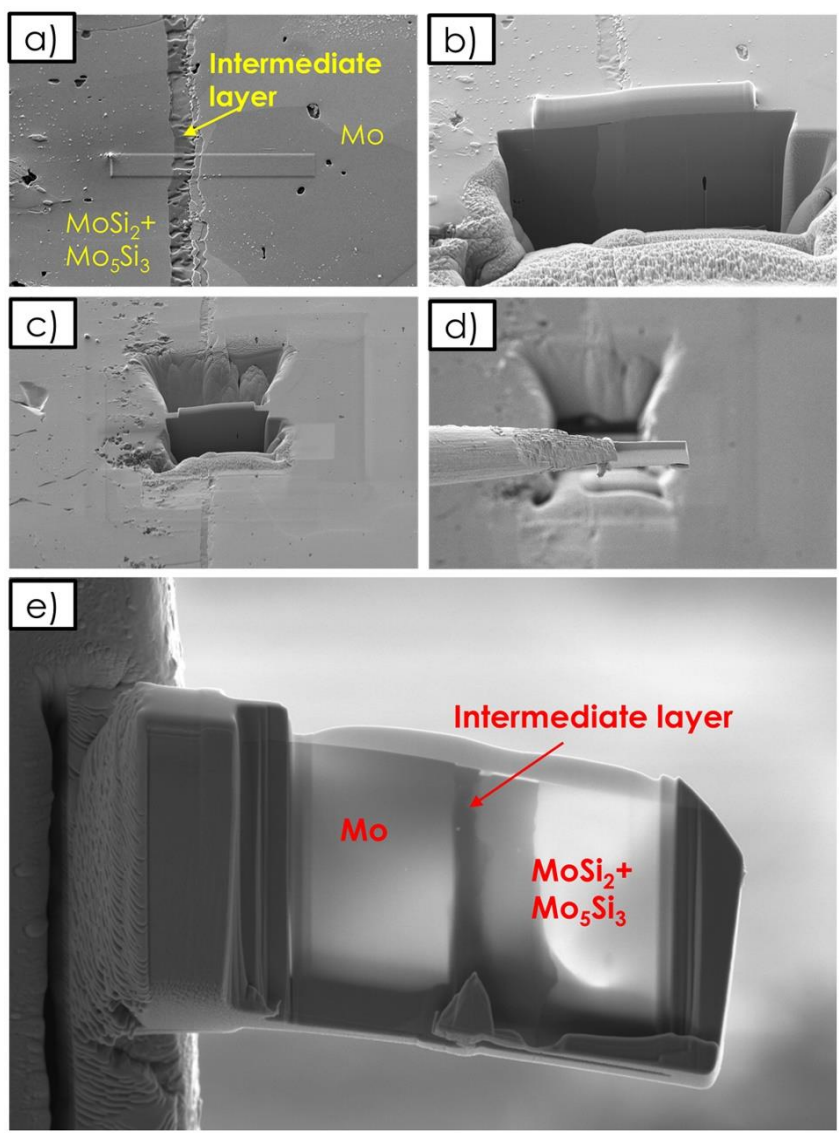

*Fig. 10. Exemplary images showing the FIB preparation of a thin foil from the Si-3.2B/Mo sample produced at 1385°C/30 min: a selection of the investigated area (a); the ion milling of a specimen (b, c); an extraction of the specimen by a manipulator (d) and an overview of the final thin foil (e).*

Figure 11a shows TEM bright field image of sample cut from the cross-sectioned Si-3.2B/Mo sample produced at 1385°C/30 min. Colored dots correspond to SAED pattern shown in Fig. 11b. Based on diffraction patterns collected from various sites of the intermediate zone (Fig. 11a, b), the intermediate layered microstructure of different Mo-based borides and borosilicide, was confirmed (Fig. 11c). Starting from the Mo substrate, first layer was composed of $Mo_2B$ phase with a thickness of about 500 nm. Next, MoB layer was identified. It was hard to determine its thickness due to area from which the sample was cut using FIB. Notably, it was the most B-rich phase formed during the liquid assisted fabrication. Lastly, the outermost phase in the interlayer was identified as $Mo_5SiB_2$ phase with a tetragonal crystal structure. The main layer was identified as tens of microns thick (Fig. 8) $MoSi_2$ phase. A bright phase located near pores (Fig. 9) was not present in the investigated TEM samples. Namely, at the Si-3.2B/Mo interface (starting from Mo substrate), as shown in Figure 14, the following compositional transition was detected: *$Mo/Mo_2B/MoB/$ $/Mo_5SiB_2/(MoSi_2+Mo_5Si_3)$*.

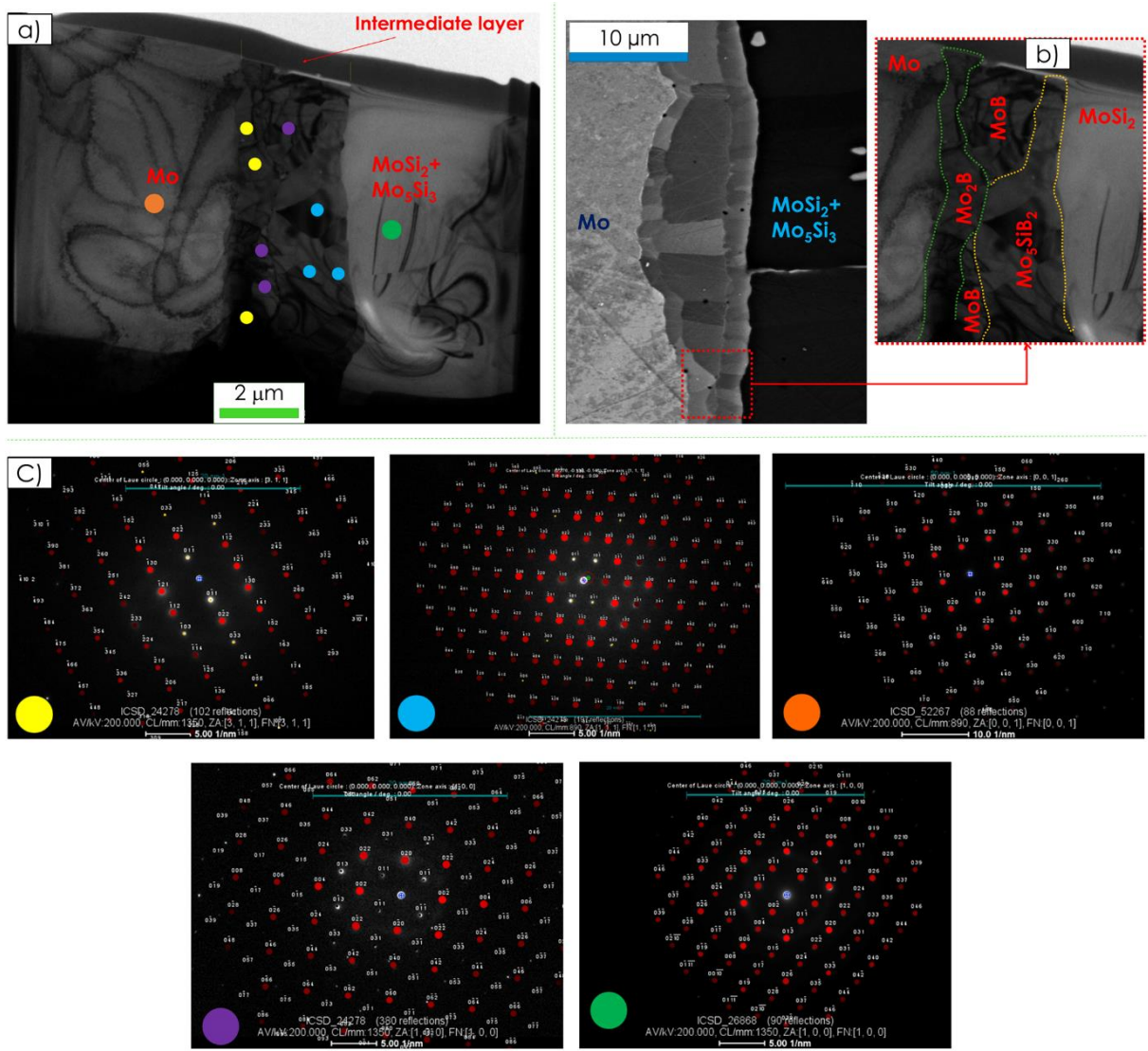

*Fig. 11. The results of TEM/SAED analyses taken from the intermediate layer of the Si-3.2B/Mo sample produced at 1385°C/30 min: a general view of the thin foil (a); a schematic representation of each phase constituent in the vicinity of intermediate layer (b); a set of collected SAED patterns (c).*

The proposed phase identification was confirmed by the obtained XRD pattern (Fig. 12) characterized by a presence of diffraction lines coming from all aforementioned phases. The existence of intermediate layer composed of a mix of B-rich compounds (MoB/Mo$_2$B/Mo$_5$SiB$_2$) are confirmed by the results of XRD analyses. It should be noted that the presence of ternary borosilicide phase was revealed at XRD spectra taken from samples tested at higher temperatures, i.e. it required greater thickness to be detected.

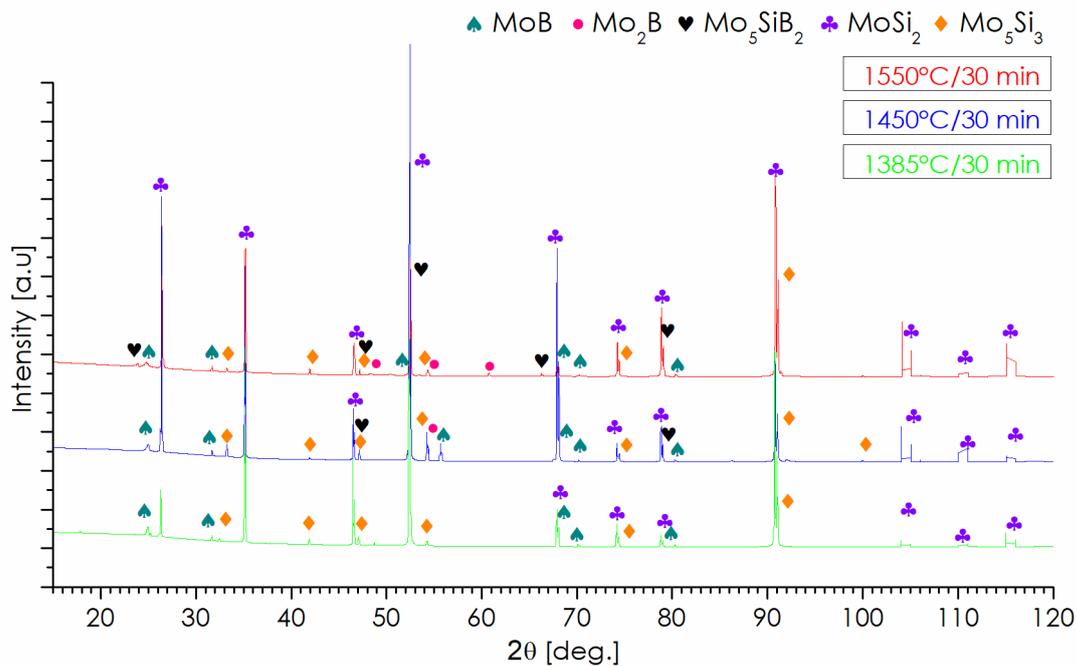

Fig. 12. A set of XRD spectra taken from the Si-3.2B/Mo sessile drop couples produced at temperatures of 1385, 1450 and 1550°C.

The as-obtained layered microstructure is typical of highly reactive mechanisms followed by precipitation processes at the interfaces. Moreover, all the intermetallic phases exhibit temperatures melting over than the testing temperature. By taking into account such outcomes, it may be concluded that the layering is driven by diffusion mechanism. Specifically, B (due to a very small atomic radius) seems to act as the element with higher diffusivity in Mo body, as compared with Si [28, 29]. For this reason, the reaction between B and Mo is faster than Mo with Si, as confirmed by the deep thermodynamic analysis provided by Yang and Chang [30]. Thus, by taking into consideration an order of the melt assisted reactions in the Si-3.2B/Mo system, it is reasonable to assume that the boride phases are produced at the beginning of interaction. A preliminary explanation for the microstructure obtained at the interlayer is that with increasing temperature and time, the diffusion of Mo and Si towards the already produced Mo borides is enhanced, resulting in the increasing of silicides thickness. To support further on such hypothesis, the precipitation of $Mo_5SiB_2$ is most probably the first step of layering which is followed by the growth of the $Mo_5Si_3$ + $MoSi_2$ two-phase layer occurring during solidification and resulting from a release of exceeding silicon after the rearrangement of $Mo_5SiB_2$ microstructure. This finding is also supported by the results of experimental works reported by Huebesch *et al.* [31] or Katrych *et al.* [24] showing that the $Mo_5SiB_2$ is produced by a peritectic reaction between the Si-rich melt and MoB compound.

The presence of a MoB+$Mo_2$B layer may even explain the similar behaviors observed during wetting experiments in terms of spreading kinetics and also in the decreasing of melting temperature. Indeed, all the binary systems with tendency to produce intermetallic compounds are energetically favored and they evolve by exothermal reactions. Such tendency might be one of the combined mechanisms occurring at the surface and producing local increase of temperature resulting in the appearance of melt phases at temperatures lower than the predicted values.

The abovementioned considerations are exclusively based on experimental observations, while thermodynamic calculations at the testing temperatures, are in progress to deeper understand on-going reactions at the interface, solidification and growth mechanisms.

## 4. Conclusions

The sessile drop experiments in the present work were intended to investigate the effect of temperature and time on the course of interfacial phenomena (wetting, spreading and reactivity) taking place between molten Si-3.2B alloy and polycrystalline Mo. The main preliminary conclusions derived from wetting tests carried out at *T*=1385, 1450 or 1550°C and at time up to 30 minutes are as follows:

a) A very good wetting and fast spreading is observed between molten Si-3.2B alloy and molybdenum substrate. Increasing of temperature from 1385 to 1550°C slightly improves the wetting kinetics.

b) The interaction is assisted by a formation of the main and the intermediate product layers. Increasing of temperature and time (from 1385 to 1550°C, and from 10 to 30 minutes) results in an increase of thickness of the main product layer (~70→120 and intermediate layer (~8→25 µm).

c) The formation of following phases is experimentally documented due to an interaction of eutectic Si-B alloy with Mo at testing conditions: ($MoSi_2$ + $Mo_5Si_3$) in the main layer; B-rich phases ($Mo_5SiB_2$/$MoB$/$MoB_2$) in the intermediate layer

Nevertheless, further thermodynamic assessment is needed to support the interpretation of experimental data and to unambiguously clarify involved reactions and growth mechanism, by calculating the phases stabilities at the testing temperatures.

From practical point of view, it should be noted that a dense and continuous intermediate layer can act as a barrier for the infiltration. As an alternative to mitigate such phenomenon in our future work we propose the replacement of bulk Mo with Mo-porous preform.

**Compliance with Ethical Standards:**

**Funding:**

The financial support given by the National Science Centre, Poland under the project no. 2018/31/N/ST8/01513 (PRELUDIUM 16) is gratefully acknowledged. The work has also received a partial financial support within the Statutory Activity of Łukasiewicz – Krakow Institute of Technology (project no: 2002/00) founded by Polish Ministry of Education and Science in 2021 year. S.L. acknowledges support of the AGH statutory project No. 16.16.110.663 and infrastructure of the International Centre of Electron Microscopy for Materials Science (IC-EM), AGH-UST. The XRD measurements were supported by the statutory research funds of the Department of Structural Materials, Military University of Technology, Warsaw, Poland (UGB22-790/2022/WAT). D.G. wishes to thank CNR-Short Term Mobility Program: call 2020 (AMMCNT – n. 0059501, 29/09/2020).

**Conflict of Interest Statement**

The authors declare that they have no conflict of interest.

The article has been submitted to Journal of Materials Science (acceptance date: 17/06/2022)page 20 of 22